\documentclass[twocolumn,preprintnumbers,amsmath,amssymb,aps,prl,superscriptaddress]{revtex4}
\usepackage{graphicx}
\usepackage{textcomp} 
\usepackage[caption=false]{subfig} 
\begin{document}
\title{Cooperative strings in glassy nanoparticles}
\author{Maxence Arutkin}
\affiliation{Laboratoire de Physico-Chimie Th\'eorique, UMR CNRS Gulliver 7083, ESPCI Paris, PSL Research University, 75005 Paris, France.}
\author{Elie Rapha\"{e}l}
\affiliation{Laboratoire de Physico-Chimie Th\'eorique, UMR CNRS Gulliver 7083, ESPCI Paris, PSL Research University, 75005 Paris, France.}
\author{James A. Forrest}
\affiliation{Laboratoire de Physico-Chimie Th\'eorique, UMR CNRS Gulliver 7083, ESPCI Paris, PSL Research University, 75005 Paris, France.}
\affiliation{Perimeter Institute for Theoretical Physics, Waterloo, Ontario N2L 2Y5, Canada.}
\affiliation{Department of Physics \& Astronomy and Guelph-Waterloo Physics Institute, University of Waterloo,  Waterloo, Ontario N2L 3G1, Canada.}
\author{Thomas Salez}
\thanks{Corresponding author: thomas.salez@espci.fr}
\affiliation{Laboratoire de Physico-Chimie Th\'eorique, UMR CNRS Gulliver 7083, ESPCI Paris, PSL Research University, 75005 Paris, France.}
\affiliation{Perimeter Institute for Theoretical Physics, Waterloo, Ontario N2L 2Y5, Canada.}
\affiliation{Global Station for Soft Matter, Global Institution for Collaborative Research and Education, Hokkaido University, Sapporo, Hokkaido 060-0808, Japan.}
\date{\today}
\begin{abstract}
Motivated by recent experimental results on glassy polymer nanoparticles, we develop a minimal theoretical framework for the glass transition in spherical confinement. This is accomplished using our cooperative-string model for supercooled dynamics, that was successful at recovering the bulk phenomenology and describing the thin-film anomalies. In particular, we obtain predictions for the mobile-layer thickness as a function of temperature, and for the effective glass-transition temperature as a function of the radius of the spherical nanoparticle -- including the existence of a critical particle radius below which vitrification never occurs. Finally, we compare the theoretical results to experimental data on polystyrene from the recent literature, and we discuss the latter.
\end{abstract}

\maketitle
\section{Introduction}
As a significant part of the ongoing research towards the understanding of the glass transition~\cite{Anderson1995,Parisi2010,Liu2010,Berthier2010,Berthier2011,Ediger2012}, the past two decades have seen significant interest in the anomalous dynamics of thin glass-forming polymer films~\cite{Mckenna2003,Ediger2014}. In particular, the observed reductions of the glass-transition temperature in thin polymer films~\cite{Keddie1994,Forrest1996,Ellison2003,Dalnoki2000,Baumchen2012} have been suggested to be either strongly influenced or caused by the enhanced dynamics in the free-surface region~\cite{Fakhraai2008,Ilton2009,Yang2010,Chai2014,Zhang2016}, and have triggered an intense theoretical activity~\cite{Ngai1998,Scheidler2000,Long2001,Herminghaus2001,Varnik2002,Baschnagel2005,Lipson2009,Forrest2013,Lam2013,Forrest2014,Mirigian2014,Hanakata2015,Salez2015}. Similarly, the effect of interfaces is expected to manifest in other geometries~\cite{Bares1975,Jackson1990}. The simplest such case is that of polymer spherical nanoparticles~\cite{Sasaki2003,Rharbi2008,Guo2011,Zhang2011,Zhang2013,Zhang2013b,Feng2013,Feng2014}, and their recent colloidal analogues~\cite{Zhang2016b}. 

To our knowledge, the first report of anomalous dynamics in polymer nanoparticles is that of Sasaki \textit{et al.}~\cite{Sasaki2003}. In that work, differential scanning calorimetry of aqueous dispersions of polystyrene nanospheres in the 21-274~nm radius range showed no evidence for a reduced glass-transition temperature, but instead revealed a radius-dependent value of the step in heat capacity at the glass transition. Analysis of the data suggested that this result is consistent with a near-surface region of size $\sim3.8$~nm not contributing to the transition. Then, Rharby used neutron scattering to measure mechanical deformations of polystyrene nanospheres individually dispersed in crosslinked polybuthylmethacrylate matrices, and deduced glass-transition temperatures that were reduced from the bulk value for nanospheres less than $\sim30$~nm in radius~\cite{Rharbi2008}. Later measurements by Zhang \textit{et al.}~\cite{Zhang2013}, and Feng \textit{et al.}~\cite{Feng2013}, showed reductions in the glass-transition temperature for spheres of larger radii, and with a strong dependence on the sphere coating. The latter fact is reminiscent of the strong effect even small amounts of residual surfactants can have on thin-film reductions in the  glass-transition temperature, as reported by Chen and Torkelson~\cite{Torkelson2016}. The above large disparity in observations may partially result from the much more difficult sample preparation in making dispersed nanospheres, as compared to thin films. Indeed, the presence of surfactants and/or residual monomers, and the uncertainty in final molecular weight could result in large variations between different experiments. Therefore, it appears necessary to establish a theoretical framework for the description of the glass transition in spherical confinement, and by this to enable comparisons between the observations in thin films and nanoparticles. As a remark, let us mention the existence of another theoretical attempt on nanoparticles, using a thermodynamical analogy between vitrification and cristallization~\cite{Zhang2001}.

In this article, we utilise the cooperative-string model~\cite{Salez2015} -- recently developed and successfully applied to thin glassy films -- for the present case of glassy nanoparticles. The general philosophy of our approach consists in combining classical free-volume and cooperativity arguments to the more recent observations of a specific string-like character of the cooperatively rearranging regions, within a minimal kinetic model allowing to address analytically and quantitatively the confinement-induced and interfacial effects. After recalling the main ingredients of the bulk description, we turn to its modification in spherical geometry and discuss the implications for experiments. In particular, we characterize the extent of the mobile-layer region as a function of temperature, and we describe the glass-transition temperature reductions as a function of particle radius, providing predictions for experiments, including the existence of a minimal radius below which vitrification never occurs. One purpose of this work is that it allows a natural separation between the effects intrinsic to glass formation, and those related to the polymeric nature of the materials for which even the thin-film geometry~\cite{Dalnoki2000} has so far defied a proper theoretical description, despite promising ideas~\cite{deGennes2000,Milner2010}.

\section{Cooperative-String Model}
In a supercooled liquid, due to crowding and caging~\cite{Gotze1998}, local rearrangements seem to require the cooperative participation of a growing number of molecules as the temperature $T$ is decreased~\cite{Gibbs1958,Adam1965}. This was related through phenomenological arguments~\cite{Doolittle1951,Edwards1986} to the vanishing of the free volume~\cite{Cohen1979} needed for relaxation, and could lead to the tremendous slowing down of glassy dynamics described by the empirical time-temperature superposition~\cite{Vogel1921,Fulcher1925,Tammann1926,Williams1955}. This relaxation process defines a temperature-dependent length scale $\xi(T)$ for the cooperatively rearranging regions, and thus for glassy dynamics in the bulk~\cite{Donth1996,Stevenson2006}. Furthermore, within some degree of polydispersity in size, numerical simulations~\cite{Donati1998,Pal2008,Betancourt2014}, and experiments~\cite{Keys2007,Zhang2011b}, suggested that those regions might take the form of unidimensional chains -- the so-called \textit{cooperative strings}. In a recent article~\cite{Salez2015}, we developed a minimal kinetic model based on those ideas, that was successful in reproducing bulk phenomenology and in describing thin-film anomalies. Below, we reproduce the main ingredients of that model for a bulk supercooled liquid.

As a preliminary, we would like to stress an important point. For the sake of simplicity, we map the real liquid state to a simple hard-sphere liquid and thus fully neglect the enthalpic contributions in the activation barriers for relaxation. Stated differently, we assume that entropic effects are dominant in the critical slowing down of supercooled liquids. This is also what the Gibbs-DiMarzio approach to glass formation in polymers would suggest, with the underlying transition essentially determined by a vanishing of the configurational entropy~\cite{Gibbs1958}. In fact, we are not trying to come up with a definitive and detailed theory of glassy dynamics, but rather address the much simpler question: what physics are we able to get out of the ideas of caging and cooperative motion? In particular, more than the bulk description already developed by Adam and Gibbs~\cite{Adam1965} or Wolynes~\cite{Stevenson2006}, our goal is to understand how such a minimal free-volume model will exhibit finite-size effects that are not always easily determined in other approaches.

Let us consider a dense assembly of small molecules with size $\lambda_{\textrm{V}}$ and average intermolecular distance $\lambda$. The volume fraction is thus $\phi\propto (\lambda_{\textrm{V}}/\lambda)^{3}$. A test molecule sits in a cage of volume $\sim \lambda^3$, with gates of length $L\sim\lambda-\lambda_{\textrm{V}}$. We set that a typical non-cooperative liquid-like local relaxation requires $L> L_{\textrm{c}}=\lambda_{\textrm{c}}-\lambda_{\textrm{V}}$, or equivalently $\lambda$ to be larger than $\lambda_{\textrm{c}}$ -- the so-called onset of cooperativity, with volume fraction $\phi_{\textrm{c}}$. Also, when $\lambda\sim\lambda_{\textrm{V}}$, the gates are completely closed ($L\sim0$) and the system is at kinetic arrest, with volume fraction $\phi_{\textrm{V}}$. For $\lambda_{\textrm{V}}<\lambda<\lambda_{\textrm{c}}$, the relaxation is possible but necessarily collective. It requires a random string-like cooperative motion involving at least $N^{*}-1$ neighbours of the test molecule, that provide a total space $(N^{*}-1)L\sim L_{\textrm{c}}-L$ by getting in close contact with each other. The test molecule thus sees a temporary larger gate, of length $L_{\textrm{c}}$, and can exit the cage. Therefore, one gets the scaling expression of the minimal number of molecules needed for a local relaxation, \textit{i. e.} the so-called cooperativity:
\begin{eqnarray}
N^{*}(\phi)=\dfrac{\left(\dfrac{\phi_{\textrm{V}}}{\phi_{\textrm{c}}}\right)^{1/3}-1}{\left(\dfrac{\phi_{\textrm{V}}}{\phi}\right)^{1/3}-1}\ . 
\label{coopN}
\end{eqnarray}
As expected, this expression reaches $1$ at the cooperative onset $\phi_{\textrm{c}}$, where solitary rearrangements are allowed, and diverges at the kinetic arrest point $\phi_{\textrm{V}}$. Note that the bulk glass-transition point $\phi_{\textrm{g}}^{\textrm{bulk}}$ lies somewhere in between those two extreme values. 

By introducing the necessity of coherence between molecular motions within a cooperative rearrangement, one showed~\cite{Salez2015} that in a typical cooperative string made of $N^*$ molecules the relaxation rate $\tau^{-1}$ follows: 
\begin{eqnarray}
\dfrac{\tau_{0}}{\tau}=\left(\dfrac{\tau_0}{\tau_{\textrm{c}}}\right)^{N^{*}}\ , 
\label{viscN}
\end{eqnarray}
where $\tau_{\textrm{c}}\sim1\ \textrm{ps}$ is a typical liquid-like relaxation time at the cooperative onset, and $\tau_0\sim10\ \textrm{fs}$ is a molecular time scale. Our description thus naturally leads to the Adam-Gibbs phenomenology~\cite{Adam1965}. While the proposed approach does globally use free volume to correlate to dynamics, it does not necessarily mean that local regions with higher free volume are always faster in a real material, despite indications that this might be true~\cite{Schoenholz2016}. In fact, the effect of coherence of molecular motion is at least as important as the free volume itself. If parts of the sample had a higher density, but also a higher coherence factor $\tau_0/\tau_{\textrm{c}}$, then those parts may relax faster. All of this is not considered in the context of the current mean-field approach, but it does mean that we do not necessarily have to assume a strict correlation between local density and local mobility.

Since, in the temperature range considered, the thermal expansion coefficient $\alpha=-(1/\phi)\textrm{d}\phi/\textrm{d}T$ of the supercooled liquid is almost constant, one has:
\begin{equation}
\phi (T)= \phi_{\textrm{V}}[1- \alpha(T-T_{\textrm{V}})]\ , 
\label{expan}
\end{equation}
where $\phi(T_{\textrm{V}})=\phi_{\textrm{V}}$ defines the Vogel temperature $T_{\textrm{V}}$. Therefore, our description naturally leads to the Vogel-Fulcher-Tammann time-temperature superposition~\cite{Vogel1921,Fulcher1925,Tammann1926} without incorporation of any enthalpic contribution, through the use of Eqs.~(\ref{coopN}),~(\ref{viscN}) and~(\ref{expan}):
\begin{eqnarray}
\label{vft}
\tau(T)= \tau_{0}\exp\left(\dfrac{A}{T-T_{\textrm{V}}}\right)\ ,
\end{eqnarray}
where $A=(T_{\textrm{c}}-T_{\textrm{V}})\ln(\tau_{\textrm{c}}/\tau_0)$, and with $\phi(T_{\textrm{c}})=\phi_{\textrm{c}}$ by definition of the onset temperature $T_{\textrm{c}}$.
\begin{figure}[t!]
\includegraphics[scale=0.22]{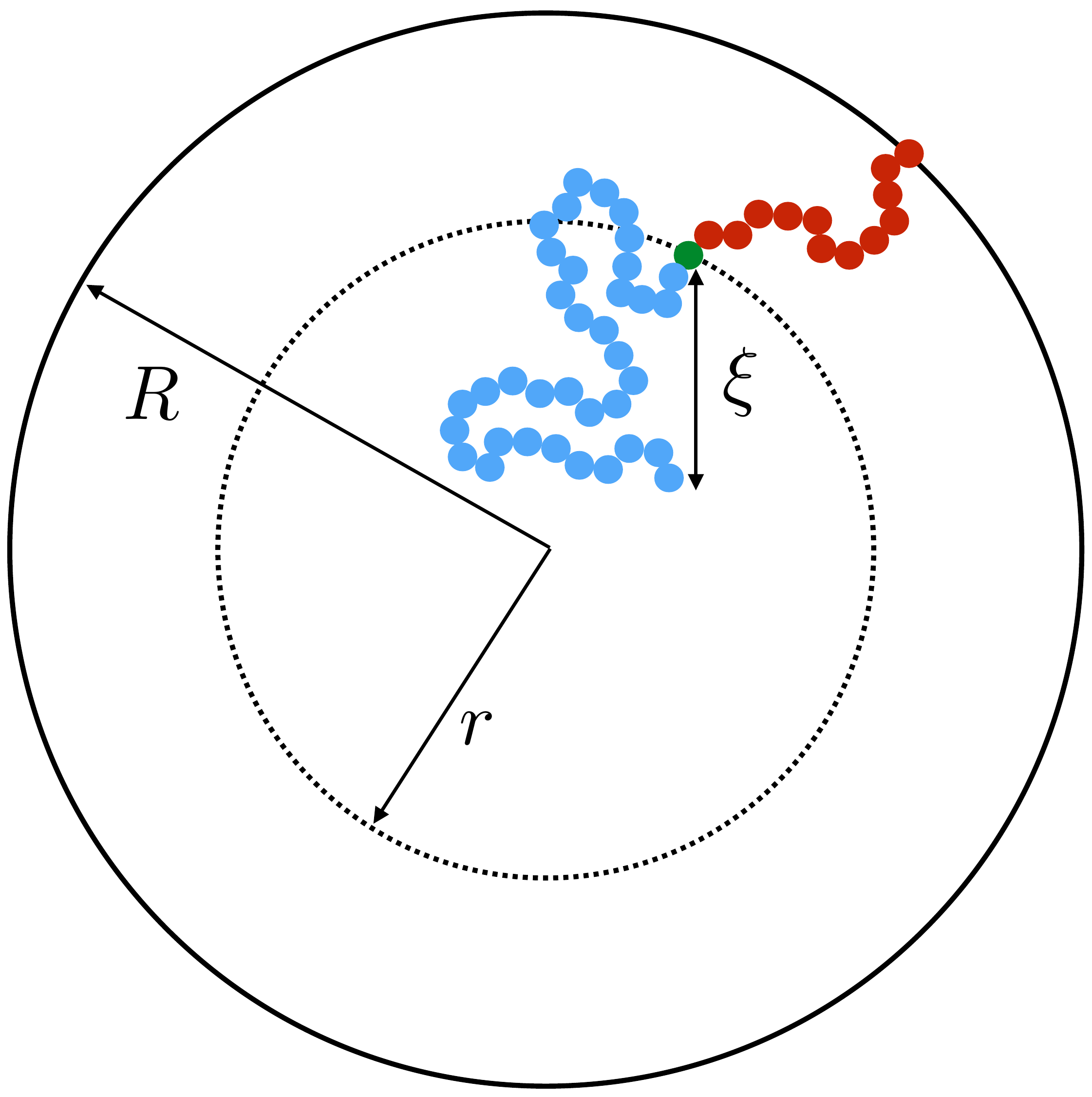}
\caption{\textit{Two string-like cooperative paths in a supercooled spherical nanoparticle of radius $R$. Relaxation of a test molecule (green) at a distance $r$ from the sphere center can occur through either a bulk cooperative string (blue) of size $\xi$ (Eq.~(\ref{coopleng})), or a truncated string (red) touching the interface.}}
\label{Fig1}
\end{figure}

As a first remark, $\tau_{\textrm{c}}/\tau_0$ should be a constant only for hard spheres. For a real liquid, the latter dimensionless relaxation time is rather expected to follow an Arrhenius law $\tau_{\textrm{c}}/\tau_0=\exp(T_{\textrm{a}}/T)$, where $k_{\textrm{B}}T_{\textrm{a}}$ is an activation energy barrier proportional to the cohesive interaction strength, and $k_{\textrm{B}}$ is the Boltzmann constant. In that case, Eq.~(\ref{vft}) would be replaced by the leading-order expression near the kinetic-arrest point:
\begin{equation}
\tau\simeq B\,\tau_0 \exp\left(\frac{C}{T-T_{\textrm{V}}}\right)\ ,
\end{equation}
where $C=T_{\textrm{a}}(T_{\textrm{c}}-T_{\textrm{V}})/T_{\textrm{V}}$ and $\log B=-C/T_{\textrm{V}}$ are two constants. Thus, the Vogel-Fulcher-Tammann form is still recovered asymptotically, which means, in this type of description, that the details of the enthalpic contributions are not essential to capture the critical slowing down in an ideal supercooled liquid. As a second remark, the Vogel-Fulcher-Tammann law is only valid over some temperature range, and in reality we do not expect the relaxation times to actually diverge (see \textit{e.~g.}~\cite{Zhao2013}). Our approach should rather be considered as a high-temperature approximation. The guiding idea behind that is to address the new spherical-confinement situation (see next section) using the most simple arguments, and in the bulk the Vogel-Fulcher-Tammann time-temperature superposition was the seminal approach before refinements: we thus went through the exact same admittedly-idealized path, before applying it to a novel geometry in order to extract its essential features. The low-temperature behaviour lies beyond the scope of the present minimal description, as it might involve the fine details of the real energy landscape.

As the bulk relaxation process presented here consists of random cooperative strings involving $N^{*}$ molecules, one can minimally describe them through ideal random walks. The length scale $\xi$ of the cooperatively rearranging regions is thus of the form $\xi\sim \lambda\sqrt{N^{*}}$, near the kinetic arrest point. Note that if we rather use more realistic self-avoiding random walks, the exponent becomes $\sim0.6$ instead of $1/2$ -- but such a refinement would be at the cost of mathematical simplicity for the confinement effects discussed below. Invoking Eqs.~(\ref{coopN}) and~(\ref{expan}), one obtains the temperature-dependent expression of the cooperativity:
\begin{eqnarray}
N^*(T)=\dfrac{T_{\textrm{c}}-T_{\textrm{V}}}{T-T_{\textrm{V}}}\ ,
\label{coop}
\end{eqnarray}
and thus an asymptotic expression for the associated length scale:
\begin{eqnarray}
\xi(T)=\lambda_{\textrm{V}}\,\sqrt{\dfrac{T_{\textrm{c}}-T_{\textrm{V}}}{T-T_{\textrm{V}}}}\ .
\label{coopleng}
\end{eqnarray}
As a consequence of this description, the cooperative length diverges at the Vogel temperature with a $-1/2$ power law, and is comparable to the molecular diameter at the cooperative onset. 

\section{Spherical confinement}
Within the context of the cooperative-string model summarized above, we now investigate the effect of confinement -- a situation that arises when the system size becomes comparable to the bulk cooperative length $\xi$. More precisely, inspired by recent experimental results~\cite{Sasaki2003,Rharbi2008,Guo2011,Zhang2011,Zhang2013,Zhang2013b,Feng2013,Feng2014}, we consider spherical nanoparticles of radius $R$. As illustrated in Fig.~\ref{Fig1}, the interface is a reservoir of free volume and as such it truncates the cooperative strings, leading to a higher local mobility. We therefore introduce the average local cooperativity $N^{*}_{\textrm{s}}(r,R,T)$, at a distance $r$ from the center of the sphere, and at temperature $T$. Intuitively, if $R-r$ is much larger than $\xi$, the interface is typically not reached with less than $N^{*}$ cooperative molecules. In contrast, when $R-r$ becomes comparable to $\xi$, the cooperative strings start to feel the interface, and the effective number of cooperative molecules needed for relaxation is reduced. Lastly, as $r$ approaches $R$, we expect $N^{*}_{\textrm{s}}$ to vanish and the relaxation to be purely liquid-like. 

Because our minimal description of the local relaxation process relies on random cooperative strings involving $N^*$ molecules, we can use a first-passage argument in the limit of large $N^*$, in order to determine $N_{\textrm{s}}^{*}$. Before doing so, we make the following remark. The typical string length in numerical simulations is rather short~\cite{Donati1998}. This is related to the fact that, due to computational time constraints, those simulations were performed at relatively high temperatures -- basically near the caging onset -- and thus address the onset of cooperative motion. In contrast, in experiments, such as the ones with vibrated granular beads~\cite{Keys2007}, or repulsive colloids~\cite{Zhang2011b} for instance, notably longer structures are seen. The clear advantage of working near the kinetic-arrest point is to get a Brownian description, and thus tractable analytical results using first-passage probability densities. This is an idealized asymptotic view valid only near the divergence point. But, as for critical phenomena in continuous phase transitions, it may allow to extract some universal features that might still be relevant away from the divergence point.

For that purpose, we define $n_{0}$ as the number of molecular units at which a given realization of a random string reaches the interface for the first ``time". If $n_{0}\geq N^{*}$, the string is bulk-like; if $n_{0}< N^{*}$, the string is truncated by the interface. Therefore, the important quantity here is the density of probability $g(t)$ of the first-passage ``time" $t=n_{0}/N^{*}$ at the interface, located at dimensionless radial position $R/\xi$, of a 3D Brownian process starting at dimensionless radial position $r/\xi$, with $r<R$. Below, we briefly summarise the main mathematical steps allowing to obtain $g(t)$ explicitly. The Laplace transform of $g(t)$ can be written as~\cite{Redner2001,Kent1978}:
\begin{equation}
\label{lt}
\hat{g}(p)=\dfrac{R}{r}\dfrac{\sinh\left(\frac{r\sqrt{2p}}{\xi}\right)}{\sinh\left(\frac{R\sqrt{2p}}{\xi}\right)}\ .
\end{equation} 
By using the Bromwich integral~\cite{Arfken1985}, one can invert the Laplace transform through:
\begin{eqnarray}
g(t)&=\dfrac{1}{2\pi i}\int\limits_{-i\infty}^{+i\infty} \textrm{ d}p\, \hat{g}(p) \exp(pt)\\
&=\sum\limits_{n=0}^{+\infty} \underset{p=p_{n}}{\textrm{Res}}[\hat{g}(p)\exp(pt)]\ ,
\label{reside}
\end{eqnarray} 
where $\underset{p=p_{n}}{\textrm{Res}}[\hat{g}(p)\exp(pt)]$ is the residue of the function $\hat{g}(p)\exp(pt)$ around the pole:
\begin{equation}
p_{n}=-n^{2}\dfrac{\pi^{2}\xi^2}{2R^{2}}\ , 
\end{equation}
with $n$ a strictly positive integer. Note that if the numerator of the right-hand side of Eq.~(\ref{lt}) vanishes at a given $p_n$, the latter is actually not a pole of $\hat{g}(p)$, and the corresponding residue equals zero which preserves the validity of Eq.~(\ref{reside}). Using l'H\^opital's rule, Eq.~(\ref{reside}) can be further reformulated into the series:
\begin{eqnarray}
g(t)=\frac{1}{uv^2}\sum\limits_{n=1}^{+\infty}(-1)^{n-1}\,\sin(ua_n)\,a_{n}\,\exp\left(-\frac{a_ n^{\,2}t}{2v^2}\right)\ ,
\end{eqnarray}
where we introduced $a_{n}=n\pi$, $u=r/R$, and $v=R/\xi$, for clarity.
\begin{figure}[t!]
\includegraphics[scale=0.25]{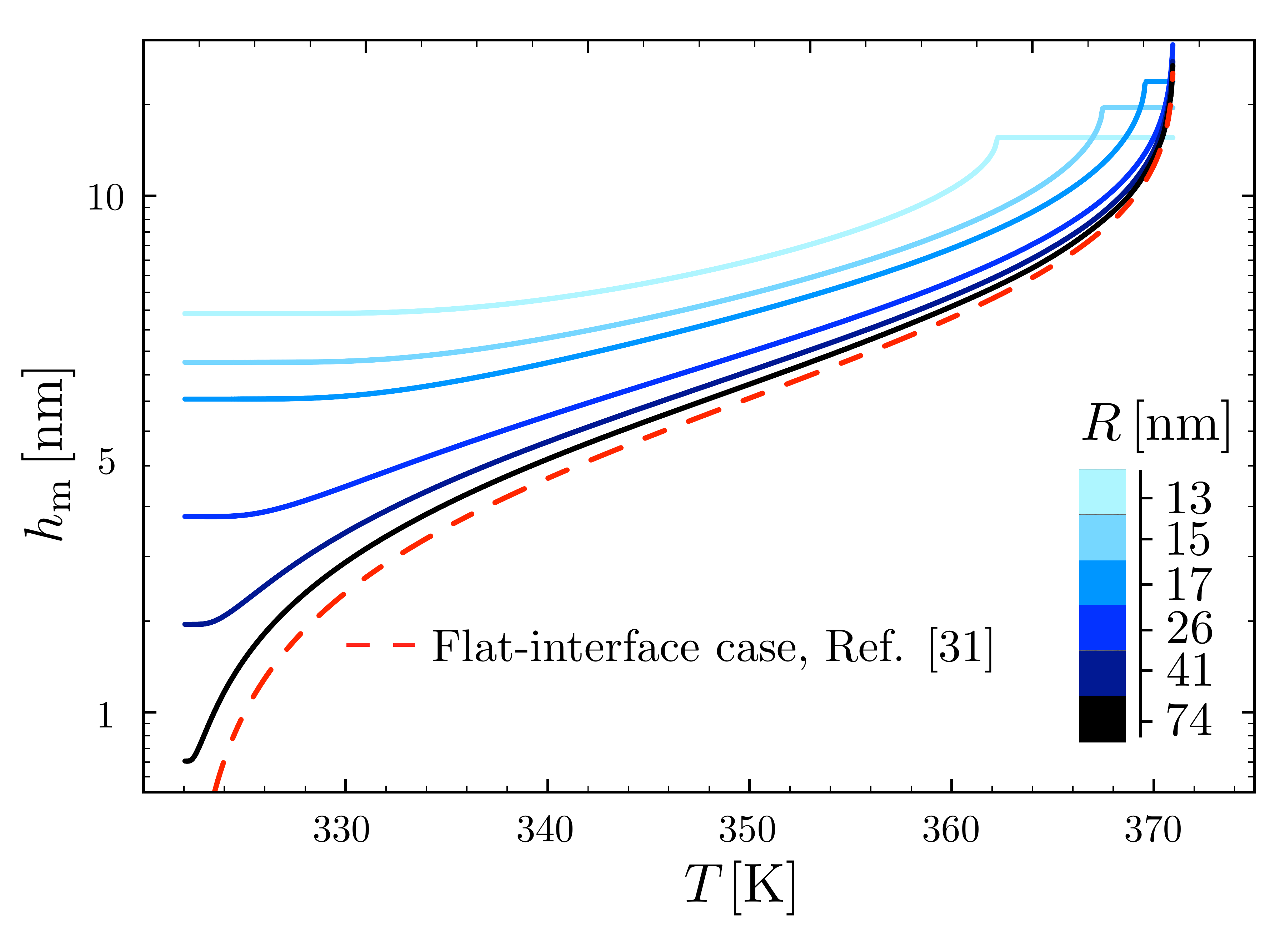}
\caption{\textit{Predicted surface mobile-layer thicknesses $h_{\textrm{m}}$ of spherical polystyrene nanoparticles as a function of temperature $T$, according to Eqs.~(\ref{fsuv}) and~(\ref{mobthick}), for different sphere radii as indicated. We used the bulk glass-transition temperature $T^{\textrm{bulk}}_{\textrm{g}} = 371\ \textrm{K}$~\cite{Rubinstein2003}, and the onset temperature $T_{\textrm{c}}=463\ \textrm{K}$~\cite{Donth1996,Kahle1997}. We fixed the molecular diameter $\lambda_{\textrm{V}}=3.7$~nm, and the Vogel temperature $T_{\textrm{V}}= 322\ \textrm{K}$, to the values previously obtained for the thin-film geometry~\cite{Salez2015}. Note that we replaced the $+\infty$ bound by $25$ in Eq.~(\ref{fsuv}), and checked that it provides sufficiently precise numerical estimates. For comparison, the dashed line indicates the flat-interface result used for the thin-film geometry~\cite{Salez2015}.}}
\label{Fig2}
\end{figure}

Knowing the first-passage probability density $g(t)$, one can now compute the average local cooperativity $N_{\textrm{s}}^{*}$ by averaging the minimum between $N^*$ and $n_0$:
\begin{eqnarray}
\label{nsuv1}
N_{\textrm{s}}^{*}(r,R,T)&=& N^*(T)\,\left\langle\textrm{min}(1,t)\right\rangle_{t}\\
&=&N^{*}(T)\,f\left(\frac{r}{R},\frac{R}{\xi(T)}\right)\ ,
\label{nsuv}
\end{eqnarray} 
where $\langle...\rangle_{t}$ indicates the average over all first-passage times. In the last equation, we introduced the truncation factor:
\begin{eqnarray}
f(u,v)=1-\int_{0}^{1}\textrm{d}t\,(1-t)\,g(t)&\quad \\
=4\dfrac{v^2}{u}\sum\limits_{n=1}^{+\infty}\dfrac{(-1)^{n-1}\sin(u a_{n})}{a_{n}^{\,3}}\left[1-\exp\left(-\frac{a_ n^{\,2}}{2v^2}\right)\right]\ ,&
\label{fsuv}
\end{eqnarray}
that takes values in the interval $\left[0,1\right]$, and which depends on the nanoparticle radius $R$, the radial location $r$ of the test molecule, and the bulk cooperative length $\xi$.

\section{Discussion}
Using the results of the previous section, one can now quantify the thickness of the mobile layer near the surface of a nanoparticle. Let us consider a temperature $T$ below the bulk glass-transition temperature $T_{\textrm{g}}^{\textrm{bulk}}$. At the interface, the relaxation is non-cooperative or liquid-like; in contrast, deep inside the sphere the relaxation is bulk-like and the sample exhibits glassy dynamics. Therefore, there must exist a distance $h_{\textrm{m}}$ from the interface, at which the local cooperativity $N_{\textrm{s}}^{*}(R-h_{\textrm{m}},R,T)$ equals the bulk cooperativity at the bulk glass transition $N^*\left({T_{\textrm{g}}^{\textrm{bulk}}}\right)$. Invoking Eqs.~(\ref{coop}),~(\ref{coopleng}), and~(\ref{nsuv}), one obtains:   
\begin{eqnarray}
f\left(\frac{R-h_{\textrm{m}}}{R},\frac{R}{\lambda_{\textrm{V}}}\sqrt{\dfrac{T-T_{\textrm{V}}}{T_{\textrm{c}}-T_{\textrm{V}}}}\right)&=&\frac{T-T_{\textrm{V}}}{T_{\textrm{g}}^{\textrm{bulk}}-T_{\textrm{V}}}\ ,
\label{mobthick}
\end{eqnarray}
which implicitly defines the temperature-dependent mobile-layer thickness $h_{\textrm{m}}(T)$. The latter is plotted in Fig.~\ref{Fig2} for several sphere radii, and with the relevant parameters for polystyrene~\cite{Rubinstein2003,Donth1996,Kahle1997,Salez2015}. A number of observations can be made. First, as expected, the curves converge at large $R$ to the flat-interface result used for the thin-film geometry~\cite{Salez2015}. Secondly, $h_{\textrm{m}}$ is an increasing function of temperature, bounded by the system size $R$. In all cases, at the bulk glass-transition temperature $T_{\textrm{g}}^{\textrm{bulk}}$, \textit{i. e.} when all the material is liquid, $h_{\textrm{m}}$ equals the system size. Thirdly, below $T_{\textrm{g}}^{\textrm{bulk}}$, $h_{\textrm{m}}$ is typically of the order of a few nanometers, consistent with the $\sim3.8$~nm estimate of Sasaki \textit{et al.} for polystyrene nanospheres~\cite{Sasaki2003}, as well as with several experimental observations on thin polystyrene films~\cite{Fakhraai2008,Ilton2009,Yang2010,Chai2014,Zhang2016}. Finally, near the Vogel temperature $T_{\textrm{V}}$, $h_{\textrm{m}}$ saturates to a finite value that can be calculated explicitly since the cooperativity $N^*$ becomes very large (see Eq.~(\ref{coop})) and thus the typical values of $t=n_0/N^*$ are much lower than 1. In that case, $f=\left\langle\textrm{min}(1,t)\right\rangle_{t}\simeq\left\langle t\right\rangle_{t}=-\hat{g}'(0)=(R^2-r^2)/(3\xi^2)$, according to Eq.~(\ref{lt}). Using this approximation, as well as Eqs.~(\ref{coopleng}) and~(\ref{mobthick}), one obtains the limiting value:
\begin{eqnarray}
\frac{h_{\textrm{m}}(T_{\textrm{V}})}{R}=1-\sqrt{1-\left(\frac{R_{\textrm{min}}}{R}\right)^2}\ ,
\end{eqnarray}
where $R_{\textrm{min}}=\sqrt{3}\, \xi(T_{\textrm{g}}^{\textrm{bulk}})\sim10$~nm, with the parameters relevant to polystyrene~\cite{Rubinstein2003,Donth1996,Kahle1997,Salez2015}. This result implies that a polystyrene nanoparticle with a nearly-free interface and a radius below a certain nanometric size remains liquid at all temperatures. Beyond its exact estimate, which might not be captured by the minimal model presented here, the existence of such a critical radius seems crucial as it might place important constraints for potential applications. On a fundamental level, the idea of a confinement-induced melting of a glass~\cite{Bares1975,Jackson1990} for a system size comparable to the bulk cooperative length $\xi(T_{\textrm{g}}^{\textrm{bulk}})$ at the bulk glass transition appears to be relevant. As a last remark, note that as $R$ diverges, $h_{\textrm{m}}(T_{\textrm{V}})$ vanishes, as in the flat-interface result used for the thin-film geometry~\cite{Salez2015}.

\begin{figure}[t!]
\includegraphics[scale=0.25]{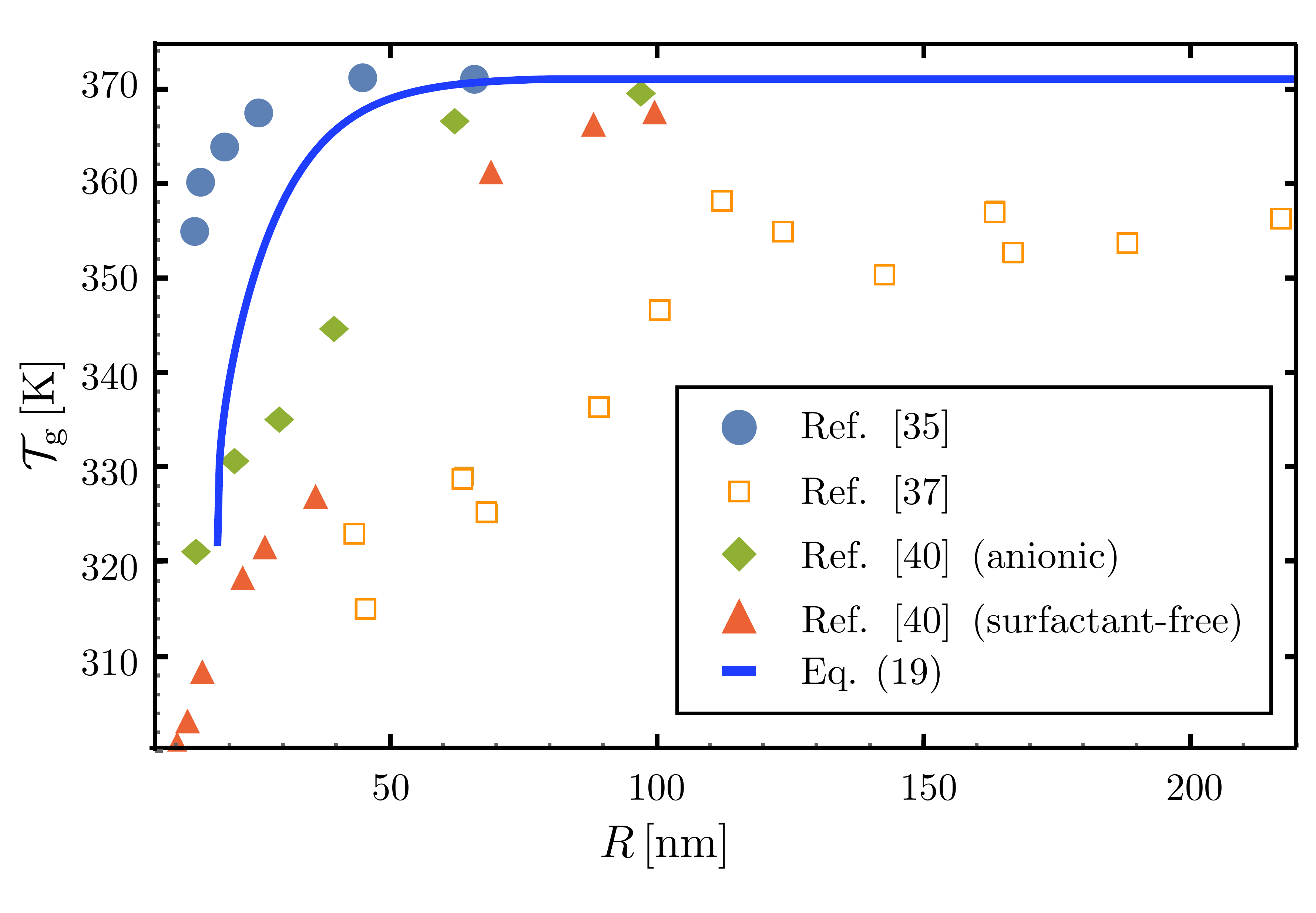}
\caption{\textit{Comparison between experimental data (symbols) for the reduced glass-transition temperature $\mathcal{T}_{\textrm{g}}(R)$ of spherical polystyrene nanoparticles~\cite{Rharbi2008,Zhang2011,Feng2013} of radius $R$, and the theory (line) given by Eq.~(\ref{tgr}) -- that invokes Eq.~(\ref{fsuv}) through $\mathcal{F}(v)=f(2^{-1/3},v)$. The fixed parameters are the bulk glass-transition temperature $T^{\textrm{bulk}}_{\textrm{g}} = 371\ \textrm{K}$~\cite{Rubinstein2003}, and the onset temperature $T_{\textrm{c}}=463\ \textrm{K}$~\cite{Donth1996,Kahle1997}. The two adjustable parameters are the molecular diameter $\lambda_{\textrm{V}}=3.7$~nm, and the Vogel temperature $T_{\textrm{V}}= 322\ \textrm{K}$, that were fixed to the values previously obtained for the thin-film geometry~\cite{Salez2015}. Note that we replaced the $+\infty$ bound by $25$ in Eq.~(\ref{fsuv}), and checked that it provides sufficiently precise numerical estimates.}}
\label{Fig3}
\end{figure}

We can now determine the effective glass-transition temperature $\mathcal{T}_{\textrm{g}}(R)$ measured in nanoparticles of radius $R$, by using the following criterion~\cite{Forrest2013,Forrest2014}: the transition occurs when half of the sample volume is liquid and the other half is glassy, \textit{i.e.} $(R-h_{\textrm{m}})/R=2^{-1/3}$ for a sphere. Introducing $\mathcal{F}(v)=f\left(2^{-1/3},v\right)$, and using Eq.~(\ref{mobthick}), we get:
\begin{eqnarray}
\label{tgr}
R\left(\mathcal{T}_{\textrm{g}} \right) = \lambda_{\textrm{V}}\,\sqrt{ \dfrac{T_{\textrm{c}}-T_{\textrm{V}}}{\mathcal{T}_{\textrm{g}}-T_{\textrm{V}}}}\,\mathcal{F}^{-1}\left( \dfrac{\mathcal{T}_{\textrm{g}}-T_{\textrm{V}}}{T_{\textrm{g}}^{\textrm{bulk}}-T_{\textrm{V}}} \right)\ ,
\end{eqnarray}
which implicitly defines $\mathcal{T}_{\textrm{g}}(R)$. The latter is plotted in Fig.~\ref{Fig3}, with the relevant parameters for polystyrene~\cite{Rubinstein2003,Donth1996,Kahle1997,Salez2015}, and it is compared to experimental results from the literature~\cite{Rharbi2008,Zhang2011,Feng2013}.
Two things become immediately obvious. On one hand, it is clear and encouraging that both the model and the data show reductions in the glass-transition temperature for spherical polystyrene nanoparticles with a radius of a few tens of nanometers. On the other hand, making more detailed comparisons is simply not possible due to large scatter in the experimental data. 

At this point, it is important to make a few remarks. First, if we shift vertically the data of Zhang \textit{et al.}~\cite{Zhang2011}, in order to enforce equal $T_{\textrm{g}}^{\textrm{bulk}}$ values between all studies, the general agreement to literature data becomes comparable to the thin-film case~\cite{Salez2015}. Secondly, the presence of residual surfactants -- a necessary ingredient in many nanosphere preparation techniques -- has been shown to have a large effect on the magnitude of the glass-transition temperature reductions in thin films~\cite{Torkelson2016}, and could thus be responsible for part of the experimental scatter. Thirdly, spherical polymer nanoparticles may well be like freestanding films, in that no part of the sample is on a substrate. It is thus fruitful here to remember the case of freestanding polystyrene thin films~\cite{Dalnoki2000}. In those samples, two different types of behaviour were observed: i) for low molecular weights, there are reductions in the glass-transition temperature that do not depend on the molecular weight, and which are similar to the ones in supported films -- well captured by our cooperative-string model~\cite{Salez2015}; ii) in contrast, for large molecular weights, the reductions in the glass-transition temperature are much more pronounced, and exhibit a dependence with molecular weight, suggesting another -- polymeric -- relaxation mode~\cite{deGennes2000,Milner2010}, still to be described theoretically in quantitative details. 

To conclude, in view of the robustness of our cooperative-string model for bulk samples and supported films~\cite{Salez2015}, and given the present results and discussion, this work could be set as a theoretical framework for describing the glassy dynamics in surfactant-free monodisperse low-molecular-weight spherical nanoparticles. Beyond the radius-dependent reduction in the glass-transition temperature, the model offers a preliminary prediction on the surface mobile-layer thickness as a function of temperature, and sets the existence of a minimal sphere radius, below which vitrification never occurs. As such, our results reveal important constraints on potential applications, and may serve as a guiding tool for future fundamental studies around the glass transition, in confinement, and at interfaces.

\section{Acknowledgments}
The authors thank Rapha\"el Voituriez, Justin Salez, Kari Dalnoki-Veress, Ulysse Mizrahi, and Nathan Israeloff for interesting discussions. They acknowledge financial support from the ESPCI Paris Sciences Chair, the Global Station for Soft Matter -- a project of Global Institution for Collaborative Research and Education at Hokkaido University -- and the Perimeter Institute for Theoretical Physics. Research at Perimeter Institute is supported by the Government of Canada through Industry Canada and by the Province of Ontario through the Ministry of Economic Development $\&$ Innovation.


\begin{thebibliography}{64}
\expandafter\ifx\csname natexlab\endcsname\relax\def\natexlab#1{#1}\fi
\expandafter\ifx\csname bibnamefont\endcsname\relax
  \def\bibnamefont#1{#1}\fi
\expandafter\ifx\csname bibfnamefont\endcsname\relax
  \def\bibfnamefont#1{#1}\fi
\expandafter\ifx\csname citenamefont\endcsname\relax
  \def\citenamefont#1{#1}\fi
\expandafter\ifx\csname url\endcsname\relax
  \def\url#1{\texttt{#1}}\fi
\expandafter\ifx\csname urlprefix\endcsname\relax\def\urlprefix{URL }\fi
\providecommand{\bibinfo}[2]{#2}
\providecommand{\eprint}[2][]{\url{#2}}

\bibitem[{\citenamefont{Anderson}(1995)}]{Anderson1995}
\bibinfo{author}{\bibfnamefont{P.~W.} \bibnamefont{Anderson}},
  \bibinfo{journal}{Science} \textbf{\bibinfo{volume}{267}},
  \bibinfo{pages}{1615} (\bibinfo{year}{1995}).

\bibitem[{\citenamefont{Parisi and Zamponi}(2010)}]{Parisi2010}
\bibinfo{author}{\bibfnamefont{G.}~\bibnamefont{Parisi}} \bibnamefont{and}
  \bibinfo{author}{\bibfnamefont{F.}~\bibnamefont{Zamponi}},
  \bibinfo{journal}{Rev. Mod. Phys.} \textbf{\bibinfo{volume}{82}},
  \bibinfo{pages}{789} (\bibinfo{year}{2010}).

\bibitem[{\citenamefont{Liu and Nagel}(2010)}]{Liu2010}
\bibinfo{author}{\bibfnamefont{A.~J.} \bibnamefont{Liu}} \bibnamefont{and}
  \bibinfo{author}{\bibfnamefont{S.~R.} \bibnamefont{Nagel}},
  \bibinfo{journal}{Annu. Rev. Cond. Mat. Phys.} \textbf{\bibinfo{volume}{1}}
  (\bibinfo{year}{2010}).

\bibitem[{\citenamefont{Berthier et~al.}(2010)\citenamefont{Berthier, Biroli,
  Bouchaud, Cipeletti, and van Saarloos}}]{Berthier2010}
\bibinfo{editor}{\bibfnamefont{L.}~\bibnamefont{Berthier}},
  \bibinfo{editor}{\bibfnamefont{G.}~\bibnamefont{Biroli}},
  \bibinfo{editor}{\bibfnamefont{J.-P.} \bibnamefont{Bouchaud}},
  \bibinfo{editor}{\bibfnamefont{L.}~\bibnamefont{Cipeletti}},
  \bibnamefont{and} \bibinfo{editor}{\bibfnamefont{W.}~\bibnamefont{van
  Saarloos}}, eds., \emph{\bibinfo{title}{Dynamical heterogeneities in glasses,
  colloids, and granular media}} (\bibinfo{publisher}{Oxford University Press},
  \bibinfo{year}{2010}).

\bibitem[{\citenamefont{Berthier and Biroli}(2011)}]{Berthier2011}
\bibinfo{author}{\bibfnamefont{L.}~\bibnamefont{Berthier}} \bibnamefont{and}
  \bibinfo{author}{\bibfnamefont{G.}~\bibnamefont{Biroli}},
  \bibinfo{journal}{Rev. Mod. Phys.} \textbf{\bibinfo{volume}{83}},
  \bibinfo{pages}{587} (\bibinfo{year}{2011}).

\bibitem[{\citenamefont{Ediger and Harrowell}(2012)}]{Ediger2012}
\bibinfo{author}{\bibfnamefont{M.~D.} \bibnamefont{Ediger}} \bibnamefont{and}
  \bibinfo{author}{\bibfnamefont{P.}~\bibnamefont{Harrowell}},
  \bibinfo{journal}{The Journal of Chemical Physics} p. \bibinfo{pages}{137}
  (\bibinfo{year}{2012}).

\bibitem[{\citenamefont{McKenna}(2003)}]{Mckenna2003}
\bibinfo{author}{\bibfnamefont{G.}~\bibnamefont{McKenna}},
  \bibinfo{journal}{The European Physical Journal E: Soft Matter and Biological
  Physics} \textbf{\bibinfo{volume}{12}}, \bibinfo{pages}{191}
  (\bibinfo{year}{2003}).

\bibitem[{\citenamefont{Ediger and Forrest}(2014)}]{Ediger2014}
\bibinfo{author}{\bibfnamefont{M.~D.} \bibnamefont{Ediger}} \bibnamefont{and}
  \bibinfo{author}{\bibfnamefont{J.~A.} \bibnamefont{Forrest}},
  \bibinfo{journal}{Macromolecules} \textbf{\bibinfo{volume}{47}},
  \bibinfo{pages}{471} (\bibinfo{year}{2014}).

\bibitem[{\citenamefont{Keddie et~al.}(1994)\citenamefont{Keddie, Jones, and
  Cory}}]{Keddie1994}
\bibinfo{author}{\bibfnamefont{J.~L.} \bibnamefont{Keddie}},
  \bibinfo{author}{\bibfnamefont{R.~A.~L.} \bibnamefont{Jones}},
  \bibnamefont{and} \bibinfo{author}{\bibfnamefont{R.~A.} \bibnamefont{Cory}},
  \bibinfo{journal}{EPL (Europhysics Letters)} \textbf{\bibinfo{volume}{27}},
  \bibinfo{pages}{59} (\bibinfo{year}{1994}).

\bibitem[{\citenamefont{Forrest et~al.}(1996)\citenamefont{Forrest,
  Dalnoki-Veress, Stevens, and Dutcher}}]{Forrest1996}
\bibinfo{author}{\bibfnamefont{J.~A.} \bibnamefont{Forrest}},
  \bibinfo{author}{\bibfnamefont{K.}~\bibnamefont{Dalnoki-Veress}},
  \bibinfo{author}{\bibfnamefont{J.~R.} \bibnamefont{Stevens}},
  \bibnamefont{and} \bibinfo{author}{\bibfnamefont{J.~R.}
  \bibnamefont{Dutcher}}, \bibinfo{journal}{Physical Review Letters}
  \textbf{\bibinfo{volume}{77}}, \bibinfo{pages}{2002} (\bibinfo{year}{1996}).

\bibitem[{\citenamefont{Ellison and Torkelson}(2003)}]{Ellison2003}
\bibinfo{author}{\bibfnamefont{C.~J.} \bibnamefont{Ellison}} \bibnamefont{and}
  \bibinfo{author}{\bibfnamefont{J.~M.} \bibnamefont{Torkelson}},
  \bibinfo{journal}{Nature Materials} \textbf{\bibinfo{volume}{2}},
  \bibinfo{pages}{695} (\bibinfo{year}{2003}).

\bibitem[{\citenamefont{{Dalnoki-Veress, K.}
  et~al.}(2000)\citenamefont{{Dalnoki-Veress, K.}, {Forrest, J. A.}, {de
  Gennes, P. G.}, and {Dutcher, J. R.}}}]{Dalnoki2000}
\bibinfo{author}{\bibnamefont{{Dalnoki-Veress, K.}}},
  \bibinfo{author}{\bibnamefont{{Forrest, J. A.}}},
  \bibinfo{author}{\bibnamefont{{de Gennes, P. G.}}}, \bibnamefont{and}
  \bibinfo{author}{\bibnamefont{{Dutcher, J. R.}}}, \bibinfo{journal}{J. Phys.
  IV France} \textbf{\bibinfo{volume}{10}}, \bibinfo{pages}{Pr7}
  (\bibinfo{year}{2000}).

\bibitem[{\citenamefont{B\"aumchen et~al.}(2012)\citenamefont{B\"aumchen,
  McGraw, Forrest, and Dalnoki-Veress}}]{Baumchen2012}
\bibinfo{author}{\bibfnamefont{O.}~\bibnamefont{B\"aumchen}},
  \bibinfo{author}{\bibfnamefont{J.~D.} \bibnamefont{McGraw}},
  \bibinfo{author}{\bibfnamefont{J.~A.} \bibnamefont{Forrest}},
  \bibnamefont{and}
  \bibinfo{author}{\bibfnamefont{K.}~\bibnamefont{Dalnoki-Veress}},
  \bibinfo{journal}{Physical Review Letters} \textbf{\bibinfo{volume}{109}},
  \bibinfo{pages}{055701} (\bibinfo{year}{2012}).

\bibitem[{\citenamefont{Fakhraai and Forrest}(2008)}]{Fakhraai2008}
\bibinfo{author}{\bibfnamefont{Z.}~\bibnamefont{Fakhraai}} \bibnamefont{and}
  \bibinfo{author}{\bibfnamefont{J.~A.} \bibnamefont{Forrest}},
  \bibinfo{journal}{Science} \textbf{\bibinfo{volume}{{\bf319}}},
  \bibinfo{pages}{600} (\bibinfo{year}{2008}).

\bibitem[{\citenamefont{Ilton et~al.}(2009)\citenamefont{Ilton, Qi, and
  Forrest}}]{Ilton2009}
\bibinfo{author}{\bibfnamefont{M.}~\bibnamefont{Ilton}},
  \bibinfo{author}{\bibfnamefont{D.}~\bibnamefont{Qi}}, \bibnamefont{and}
  \bibinfo{author}{\bibfnamefont{J.~A.} \bibnamefont{Forrest}},
  \bibinfo{journal}{Macromolecules} \textbf{\bibinfo{volume}{42}},
  \bibinfo{pages}{6851} (\bibinfo{year}{2009}).

\bibitem[{\citenamefont{Yang et~al.}(2010)\citenamefont{Yang, Fujii, Lee, Lam,
  and Tsui}}]{Yang2010}
\bibinfo{author}{\bibfnamefont{Z.}~\bibnamefont{Yang}},
  \bibinfo{author}{\bibfnamefont{Y.}~\bibnamefont{Fujii}},
  \bibinfo{author}{\bibfnamefont{F.~K.} \bibnamefont{Lee}},
  \bibinfo{author}{\bibfnamefont{C.-H.} \bibnamefont{Lam}}, \bibnamefont{and}
  \bibinfo{author}{\bibfnamefont{O.~K.~C.} \bibnamefont{Tsui}},
  \bibinfo{journal}{Science} \textbf{\bibinfo{volume}{328}},
  \bibinfo{pages}{1676} (\bibinfo{year}{2010}).

\bibitem[{\citenamefont{Chai et~al.}(2014)\citenamefont{Chai, Salez, McGraw,
  Benzaquen, Dalnoki-Veress, Rapha\"el, and Forrest}}]{Chai2014}
\bibinfo{author}{\bibfnamefont{Y.}~\bibnamefont{Chai}},
  \bibinfo{author}{\bibfnamefont{T.}~\bibnamefont{Salez}},
  \bibinfo{author}{\bibfnamefont{J.~D.} \bibnamefont{McGraw}},
  \bibinfo{author}{\bibfnamefont{M.}~\bibnamefont{Benzaquen}},
  \bibinfo{author}{\bibfnamefont{K.}~\bibnamefont{Dalnoki-Veress}},
  \bibinfo{author}{\bibfnamefont{E.}~\bibnamefont{Rapha\"el}},
  \bibnamefont{and} \bibinfo{author}{\bibfnamefont{J.~A.}
  \bibnamefont{Forrest}}, \bibinfo{journal}{Science}
  \textbf{\bibinfo{volume}{343}}, \bibinfo{pages}{994} (\bibinfo{year}{2014}).

\bibitem[{\citenamefont{Zhang and Yu}(2016)}]{Zhang2016}
\bibinfo{author}{\bibfnamefont{W.}~\bibnamefont{Zhang}} \bibnamefont{and}
  \bibinfo{author}{\bibfnamefont{L.}~\bibnamefont{Yu}},
  \bibinfo{journal}{Macromolecules} \textbf{\bibinfo{volume}{49}},
  \bibinfo{pages}{731} (\bibinfo{year}{2016}).

\bibitem[{\citenamefont{Ngai et~al.}(1998)\citenamefont{Ngai, Rizos, and
  Plazek}}]{Ngai1998}
\bibinfo{author}{\bibfnamefont{K.}~\bibnamefont{Ngai}},
  \bibinfo{author}{\bibfnamefont{A.}~\bibnamefont{Rizos}}, \bibnamefont{and}
  \bibinfo{author}{\bibfnamefont{D.}~\bibnamefont{Plazek}},
  \bibinfo{journal}{Journal of Non-Crystalline Solids}
  \textbf{\bibinfo{volume}{235-237}}, \bibinfo{pages}{435}
  (\bibinfo{year}{1998}).

\bibitem[{\citenamefont{Scheidler et~al.}(2000)\citenamefont{Scheidler, Kob,
  and Binder}}]{Scheidler2000}
\bibinfo{author}{\bibfnamefont{P.}~\bibnamefont{Scheidler}},
  \bibinfo{author}{\bibfnamefont{W.}~\bibnamefont{Kob}}, \bibnamefont{and}
  \bibinfo{author}{\bibfnamefont{K.}~\bibnamefont{Binder}},
  \bibinfo{journal}{EPL (Europhysics Letters)} \textbf{\bibinfo{volume}{52}},
  \bibinfo{pages}{277} (\bibinfo{year}{2000}).

\bibitem[{\citenamefont{Long and Lequeux}(2001)}]{Long2001}
\bibinfo{author}{\bibfnamefont{D.}~\bibnamefont{Long}} \bibnamefont{and}
  \bibinfo{author}{\bibfnamefont{F.}~\bibnamefont{Lequeux}},
  \bibinfo{journal}{European Physical Journal E} \textbf{\bibinfo{volume}{4}},
  \bibinfo{pages}{371} (\bibinfo{year}{2001}).

\bibitem[{\citenamefont{Herminghaus et~al.}(2001)\citenamefont{Herminghaus,
  Jacobs, and Seemann}}]{Herminghaus2001}
\bibinfo{author}{\bibfnamefont{S.}~\bibnamefont{Herminghaus}},
  \bibinfo{author}{\bibfnamefont{K.}~\bibnamefont{Jacobs}}, \bibnamefont{and}
  \bibinfo{author}{\bibfnamefont{R.}~\bibnamefont{Seemann}},
  \bibinfo{journal}{European Physical Journal E} \textbf{\bibinfo{volume}{5}},
  \bibinfo{pages}{531} (\bibinfo{year}{2001}).

\bibitem[{\citenamefont{Varnik et~al.}(2002)\citenamefont{Varnik, Baschnagel,
  and Binder}}]{Varnik2002}
\bibinfo{author}{\bibfnamefont{F.}~\bibnamefont{Varnik}},
  \bibinfo{author}{\bibfnamefont{J.}~\bibnamefont{Baschnagel}},
  \bibnamefont{and} \bibinfo{author}{\bibfnamefont{K.}~\bibnamefont{Binder}},
  \bibinfo{journal}{Phys. Rev. E} \textbf{\bibinfo{volume}{65}},
  \bibinfo{pages}{021507} (\bibinfo{year}{2002}).

\bibitem[{\citenamefont{Baschnagel and Varnik}(2005)}]{Baschnagel2005}
\bibinfo{author}{\bibfnamefont{J.}~\bibnamefont{Baschnagel}} \bibnamefont{and}
  \bibinfo{author}{\bibfnamefont{F.}~\bibnamefont{Varnik}},
  \bibinfo{journal}{Journal of Physics: Condensed Matter}
  \textbf{\bibinfo{volume}{17}}, \bibinfo{pages}{R851} (\bibinfo{year}{2005}).

\bibitem[{\citenamefont{Lipson and Milner}(2009)}]{Lipson2009}
\bibinfo{author}{\bibfnamefont{J.~E.~G.} \bibnamefont{Lipson}}
  \bibnamefont{and} \bibinfo{author}{\bibfnamefont{S.~T.}
  \bibnamefont{Milner}}, \bibinfo{journal}{European Physical Journal B}
  \textbf{\bibinfo{volume}{72}}, \bibinfo{pages}{133} (\bibinfo{year}{2009}).

\bibitem[{\citenamefont{Forrest}(2013)}]{Forrest2013}
\bibinfo{author}{\bibfnamefont{J.~A.} \bibnamefont{Forrest}},
  \bibinfo{journal}{The Journal of Chemical Physics}
  \textbf{\bibinfo{volume}{139}}, \bibinfo{pages}{084702}
  (\bibinfo{year}{2013}).

\bibitem[{\citenamefont{Lam and Tsui}(2013)}]{Lam2013}
\bibinfo{author}{\bibfnamefont{C.-H.} \bibnamefont{Lam}} \bibnamefont{and}
  \bibinfo{author}{\bibfnamefont{O.~K.~C.} \bibnamefont{Tsui}},
  \bibinfo{journal}{Phys. Rev. E} \textbf{\bibinfo{volume}{88}},
  \bibinfo{pages}{042604} (\bibinfo{year}{2013}).

\bibitem[{\citenamefont{Forrest and Dalnoki-Veress}(2014)}]{Forrest2014}
\bibinfo{author}{\bibfnamefont{J.~A.} \bibnamefont{Forrest}} \bibnamefont{and}
  \bibinfo{author}{\bibfnamefont{K.}~\bibnamefont{Dalnoki-Veress}},
  \bibinfo{journal}{ACS Macro Letters} \textbf{\bibinfo{volume}{3}},
  \bibinfo{pages}{310} (\bibinfo{year}{2014}).

\bibitem[{\citenamefont{Mirigian and Schweizer}(2014)}]{Mirigian2014}
\bibinfo{author}{\bibfnamefont{S.}~\bibnamefont{Mirigian}} \bibnamefont{and}
  \bibinfo{author}{\bibfnamefont{K.~S.} \bibnamefont{Schweizer}},
  \bibinfo{journal}{The Journal of Chemical Physics}
  \textbf{\bibinfo{volume}{141}}, \bibinfo{pages}{161103}
  (\bibinfo{year}{2014}).
  
  \bibitem[{\citenamefont{Hanakata et~al.}(2015)\citenamefont{Hanakata, Pazmino~Betancourt, Douglas, and Starr}}]{Hanakata2015}
\bibinfo{author}{\bibfnamefont{P.~Z.}~\bibnamefont{Hanakata}},
  \bibinfo{author}{\bibfnamefont{B.~A.}~\bibnamefont{Pazmino~Betancourt}},
  \bibinfo{author}{\bibfnamefont{J.~F.}~\bibnamefont{Douglas}},
  \bibnamefont{and} \bibinfo{author}{\bibfnamefont{F.~W.}
  \bibnamefont{Starr}}, \bibinfo{journal}{The Journal of Chemical Physics} \textbf{\bibinfo{volume}{142}}, \bibinfo{pages}{234907}
  (\bibinfo{year}{2015}).


\bibitem[{\citenamefont{Salez et~al.}(2015)\citenamefont{Salez, Salez,
  Dalnoki-Veress, Rapha\"el, and Forrest}}]{Salez2015}
\bibinfo{author}{\bibfnamefont{T.}~\bibnamefont{Salez}},
  \bibinfo{author}{\bibfnamefont{J.}~\bibnamefont{Salez}},
  \bibinfo{author}{\bibfnamefont{K.}~\bibnamefont{Dalnoki-Veress}},
  \bibinfo{author}{\bibfnamefont{E.}~\bibnamefont{Rapha\"el}},
  \bibnamefont{and} \bibinfo{author}{\bibfnamefont{J.~A.}
  \bibnamefont{Forrest}}, \bibinfo{journal}{Proceedings of the National Academy
  of Sciences of the USA} \textbf{\bibinfo{volume}{112}}, \bibinfo{pages}{8227}
  (\bibinfo{year}{2015}).

\bibitem[{\citenamefont{Bares}(1975)}]{Bares1975}
\bibinfo{author}{\bibfnamefont{J.} \bibnamefont{Bares}},
  \bibinfo{journal}{Macrocmolecules}
  \textbf{\bibinfo{volume}{8}}, \bibinfo{pages}{244} (\bibinfo{year}{1975}).
  
\bibitem[{\citenamefont{Jackson and McKenna}(1990)}]{Jackson1990}
\bibinfo{author}{\bibfnamefont{C.~L.} \bibnamefont{Jackson}} \bibnamefont{and}
  \bibinfo{author}{\bibfnamefont{G.~B.} \bibnamefont{McKenna}},
  \bibinfo{journal}{The Journal of Chemical Physics}
  \textbf{\bibinfo{volume}{93}}, \bibinfo{pages}{9002} (\bibinfo{year}{1990}).

\bibitem[{\citenamefont{Sasaki et~al.}(2003)\citenamefont{Sasaki, Shimizu,
  Mourey, Thurau, and Ediger}}]{Sasaki2003}
\bibinfo{author}{\bibfnamefont{T.}~\bibnamefont{Sasaki}},
  \bibinfo{author}{\bibfnamefont{A.}~\bibnamefont{Shimizu}},
  \bibinfo{author}{\bibfnamefont{T.~H.} \bibnamefont{Mourey}},
  \bibinfo{author}{\bibfnamefont{C.~T.} \bibnamefont{Thurau}},
  \bibnamefont{and} \bibinfo{author}{\bibfnamefont{M.~D.}
  \bibnamefont{Ediger}}, \bibinfo{journal}{The Journal of Chemical Physics}
  \textbf{\bibinfo{volume}{119}}, \bibinfo{pages}{8730} (\bibinfo{year}{2003}).

\bibitem[{\citenamefont{Rharbi}(2008)}]{Rharbi2008}
\bibinfo{author}{\bibfnamefont{Y.}~\bibnamefont{Rharbi}},
  \bibinfo{journal}{Phys. Rev. E} \textbf{\bibinfo{volume}{77}},
  \bibinfo{pages}{031806} (\bibinfo{year}{2008}).

\bibitem[{\citenamefont{Guo et~al.}(2011)\citenamefont{Guo, Zhang, Lai,
  Priestley, D'Acunzi, and Fytas}}]{Guo2011}
\bibinfo{author}{\bibfnamefont{Y.}~\bibnamefont{Guo}},
  \bibinfo{author}{\bibfnamefont{C.}~\bibnamefont{Zhang}},
  \bibinfo{author}{\bibfnamefont{C.}~\bibnamefont{Lai}},
  \bibinfo{author}{\bibfnamefont{R.~D.} \bibnamefont{Priestley}},
  \bibinfo{author}{\bibfnamefont{M.}~\bibnamefont{D'Acunzi}}, \bibnamefont{and}
  \bibinfo{author}{\bibfnamefont{G.}~\bibnamefont{Fytas}},
  \bibinfo{journal}{ACS Nano} \textbf{\bibinfo{volume}{5}},
  \bibinfo{pages}{5365} (\bibinfo{year}{2011}).

\bibitem[{\citenamefont{Zhang et~al.}(2011)\citenamefont{Zhang, Guo, and
  Priestley}}]{Zhang2011}
\bibinfo{author}{\bibfnamefont{C.}~\bibnamefont{Zhang}},
  \bibinfo{author}{\bibfnamefont{Y.}~\bibnamefont{Guo}}, \bibnamefont{and}
  \bibinfo{author}{\bibfnamefont{R.~D.} \bibnamefont{Priestley}},
  \bibinfo{journal}{Macromolecules} \textbf{\bibinfo{volume}{44}},
  \bibinfo{pages}{4001} (\bibinfo{year}{2011}).

\bibitem[{\citenamefont{Zhang et~al.}(2013{\natexlab{a}})\citenamefont{Zhang,
  Guo, Shepard, and Priestley}}]{Zhang2013}
\bibinfo{author}{\bibfnamefont{C.}~\bibnamefont{Zhang}},
  \bibinfo{author}{\bibfnamefont{Y.}~\bibnamefont{Guo}},
  \bibinfo{author}{\bibfnamefont{K.~B.} \bibnamefont{Shepard}},
  \bibnamefont{and} \bibinfo{author}{\bibfnamefont{R.~D.}
  \bibnamefont{Priestley}}, \bibinfo{journal}{The Journal of Physical Chemistry
  Letters} \textbf{\bibinfo{volume}{4}}, \bibinfo{pages}{431}
  (\bibinfo{year}{2013}{\natexlab{a}}).

\bibitem[{\citenamefont{Zhang et~al.}(2013{\natexlab{b}})\citenamefont{Zhang,
  Boucher, Cangialosi, and Priestley}}]{Zhang2013b}
\bibinfo{author}{\bibfnamefont{C.}~\bibnamefont{Zhang}},
  \bibinfo{author}{\bibfnamefont{V.~M.} \bibnamefont{Boucher}},
  \bibinfo{author}{\bibfnamefont{D.}~\bibnamefont{Cangialosi}},
  \bibnamefont{and} \bibinfo{author}{\bibfnamefont{R.~D.}
  \bibnamefont{Priestley}}, \bibinfo{journal}{Polymer}
  \textbf{\bibinfo{volume}{54}}, \bibinfo{pages}{230}
  (\bibinfo{year}{2013}{\natexlab{b}}).

\bibitem[{\citenamefont{Feng et~al.}(2013)\citenamefont{Feng, Li, Liu, Mai, Wu,
  Liang, Gaoab, and Zhu}}]{Feng2013}
\bibinfo{author}{\bibfnamefont{S.}~\bibnamefont{Feng}},
  \bibinfo{author}{\bibfnamefont{Z.~Y.} \bibnamefont{Li}},
  \bibinfo{author}{\bibfnamefont{R.}~\bibnamefont{Liu}},
  \bibinfo{author}{\bibfnamefont{B.~Y.} \bibnamefont{Mai}},
  \bibinfo{author}{\bibfnamefont{Q.}~\bibnamefont{Wu}},
  \bibinfo{author}{\bibfnamefont{G.~D.} \bibnamefont{Liang}},
  \bibinfo{author}{\bibfnamefont{H.~Y.} \bibnamefont{Gaoab}}, \bibnamefont{and}
  \bibinfo{author}{\bibfnamefont{F.~M.} \bibnamefont{Zhu}},
  \bibinfo{journal}{Soft Matter} \textbf{\bibinfo{volume}{9}},
  \bibinfo{pages}{4614} (\bibinfo{year}{2013}).

\bibitem[{\citenamefont{Feng et~al.}(2014)\citenamefont{Feng, Chen, Mai, Wei,
  Zheng, Wu, Liang, Gaoab, and Zhu}}]{Feng2014}
\bibinfo{author}{\bibfnamefont{S.}~\bibnamefont{Feng}},
  \bibinfo{author}{\bibfnamefont{Y.}~\bibnamefont{Chen}},
  \bibinfo{author}{\bibfnamefont{B.}~\bibnamefont{Mai}},
  \bibinfo{author}{\bibfnamefont{W.}~\bibnamefont{Wei}},
  \bibinfo{author}{\bibfnamefont{C.}~\bibnamefont{Zheng}},
  \bibinfo{author}{\bibfnamefont{Q.}~\bibnamefont{Wu}},
  \bibinfo{author}{\bibfnamefont{G.~D.} \bibnamefont{Liang}},
  \bibinfo{author}{\bibfnamefont{H.~Y.} \bibnamefont{Gaoab}}, \bibnamefont{and}
  \bibinfo{author}{\bibfnamefont{F.~M.} \bibnamefont{Zhu}},
  \bibinfo{journal}{Phys. Chem. Chem. Phys.} \textbf{\bibinfo{volume}{16}},
  \bibinfo{pages}{15941} (\bibinfo{year}{2014}).

\bibitem[{\citenamefont{Zhang and Cheng}(2016)}]{Zhang2016b}
\bibinfo{author}{\bibfnamefont{B.}~\bibnamefont{Zhang}} \bibnamefont{and}
  \bibinfo{author}{\bibfnamefont{X.}~\bibnamefont{Cheng}},
  \bibinfo{journal}{Phys. Rev. Lett.} \textbf{\bibinfo{volume}{116}},
  \bibinfo{pages}{098302} (\bibinfo{year}{2016}).

\bibitem[{\citenamefont{Chen and Torkelson}(2016)}]{Torkelson2016}
\bibinfo{author}{\bibfnamefont{L.}~\bibnamefont{Chen}} \bibnamefont{and}
  \bibinfo{author}{\bibfnamefont{J.~M.} \bibnamefont{Torkelson}},
  \bibinfo{journal}{Polymer} \textbf{\bibinfo{volume}{86}},
  \bibinfo{pages}{226} (\bibinfo{year}{2016}).
  
  \bibitem[{\citenamefont{Zhang}(2001)}]{Zhang2001}
  \bibinfo{author}{\bibfnamefont{Z.}~\bibnamefont{Zhang}},
\bibinfo{author}{\bibfnamefont{M.}~\bibnamefont{Zhao}}, \bibnamefont{and}
  \bibinfo{author}{\bibfnamefont{Q.} \bibnamefont{Jiang}},
  \bibinfo{journal}{Physica B} \textbf{\bibinfo{volume}{293}},
  \bibinfo{pages}{232} (\bibinfo{year}{2001}).

\bibitem[{\citenamefont{de~Gennes}(2000)}]{deGennes2000}
\bibinfo{author}{\bibfnamefont{P.-G.}~\bibnamefont{de~Gennes}},
  \bibinfo{journal}{The European Physical Journal E}
  \textbf{\bibinfo{volume}{2}}, \bibinfo{pages}{201} (\bibinfo{year}{2000}),
  ISSN \bibinfo{issn}{1292-8941}.

\bibitem[{\citenamefont{Milner and Lipson}(2010)}]{Milner2010}
\bibinfo{author}{\bibfnamefont{S.~T.} \bibnamefont{Milner}} \bibnamefont{and}
  \bibinfo{author}{\bibfnamefont{J.~E.~G.} \bibnamefont{Lipson}},
  \bibinfo{journal}{Macromolecules} \textbf{\bibinfo{volume}{9865}},
  \bibinfo{pages}{43} (\bibinfo{year}{2010}).

\bibitem[{\citenamefont{G\"otze}(1998)}]{Gotze1998}
\bibinfo{author}{\bibfnamefont{W.}~\bibnamefont{G\"otze}},
  \bibinfo{journal}{Condensed Matter Physics}
  \textbf{\bibinfo{volume}{{\bf1}}}, \bibinfo{pages}{873}
  (\bibinfo{year}{1998}).

\bibitem[{\citenamefont{Gibbs and DiMarzio}(1958)}]{Gibbs1958}
\bibinfo{author}{\bibfnamefont{J.~H.} \bibnamefont{Gibbs}} \bibnamefont{and}
  \bibinfo{author}{\bibfnamefont{E.~A.} \bibnamefont{DiMarzio}},
  \bibinfo{journal}{The Journal of Chemical Physics}
  \textbf{\bibinfo{volume}{28}}, \bibinfo{pages}{373} (\bibinfo{year}{1958}).

\bibitem[{\citenamefont{Adam and Gibbs}(1965)}]{Adam1965}
\bibinfo{author}{\bibfnamefont{G.}~\bibnamefont{Adam}} \bibnamefont{and}
  \bibinfo{author}{\bibfnamefont{J.~H.} \bibnamefont{Gibbs}},
  \bibinfo{journal}{The Journal of Chemical Physics}
  \textbf{\bibinfo{volume}{43}}, \bibinfo{pages}{139} (\bibinfo{year}{1965}).

\bibitem[{\citenamefont{Doolittle}(1951)}]{Doolittle1951}
\bibinfo{author}{\bibfnamefont{A.~K.} \bibnamefont{Doolittle}},
  \bibinfo{journal}{Journal of Applied Physics} \textbf{\bibinfo{volume}{22}},
  \bibinfo{pages}{1471} (\bibinfo{year}{1951}).

\bibitem[{\citenamefont{Edwards and Vilgis}(1986)}]{Edwards1986}
\bibinfo{author}{\bibfnamefont{S.~F.} \bibnamefont{Edwards}} \bibnamefont{and}
  \bibinfo{author}{\bibfnamefont{T.}~\bibnamefont{Vilgis}},
  \bibinfo{journal}{Physica Scripta} \textbf{\bibinfo{volume}{T13}},
  \bibinfo{pages}{7} (\bibinfo{year}{1986}).

\bibitem[{\citenamefont{Cohen and Grest}(1979)}]{Cohen1979}
\bibinfo{author}{\bibfnamefont{M.~H.} \bibnamefont{Cohen}} \bibnamefont{and}
  \bibinfo{author}{\bibfnamefont{G.~S.} \bibnamefont{Grest}},
  \bibinfo{journal}{Phys. Rev. B} \textbf{\bibinfo{volume}{20}},
  \bibinfo{pages}{1077} (\bibinfo{year}{1979}).

\bibitem[{\citenamefont{Vogel}(1921)}]{Vogel1921}
\bibinfo{author}{\bibfnamefont{H.}~\bibnamefont{Vogel}},
  \bibinfo{journal}{Physikalische Zeitschrift} \textbf{\bibinfo{volume}{22}},
  \bibinfo{pages}{645} (\bibinfo{year}{1921}).

\bibitem[{\citenamefont{Fulcher}(1925)}]{Fulcher1925}
\bibinfo{author}{\bibfnamefont{G.~S.} \bibnamefont{Fulcher}},
  \bibinfo{journal}{Journal of the American Ceramic Society}
  \textbf{\bibinfo{volume}{8}}, \bibinfo{pages}{339} (\bibinfo{year}{1925}).

\bibitem[{\citenamefont{Tammann and Hesse}(1926)}]{Tammann1926}
\bibinfo{author}{\bibfnamefont{G.}~\bibnamefont{Tammann}} \bibnamefont{and}
  \bibinfo{author}{\bibfnamefont{W.}~\bibnamefont{Hesse}},
  \bibinfo{journal}{Zeitschrift für Anorganische und Allgemeine Chemie}
  \textbf{\bibinfo{volume}{156}}, \bibinfo{pages}{245} (\bibinfo{year}{1926}).

\bibitem[{\citenamefont{Williams et~al.}(1955)\citenamefont{Williams, Landel,
  and Ferry}}]{Williams1955}
\bibinfo{author}{\bibfnamefont{M.~L.} \bibnamefont{Williams}},
  \bibinfo{author}{\bibfnamefont{R.~F.} \bibnamefont{Landel}},
  \bibnamefont{and} \bibinfo{author}{\bibfnamefont{J.~D.} \bibnamefont{Ferry}},
  \bibinfo{journal}{Journal of the American Chemical Society}
  \textbf{\bibinfo{volume}{{\bf 77}}}, \bibinfo{pages}{3701}
  (\bibinfo{year}{1955}).

\bibitem[{\citenamefont{Donth}(1996)}]{Donth1996}
\bibinfo{author}{\bibfnamefont{E.}~\bibnamefont{Donth}},
  \bibinfo{journal}{Journal of Polymer Science: Part B: Polymer Physics}
  \textbf{\bibinfo{volume}{34}}, \bibinfo{pages}{2881} (\bibinfo{year}{1996}).

\bibitem[{\citenamefont{Stevenson et~al.}(2006)\citenamefont{Stevenson,
  Schmalian, and Wolynes}}]{Stevenson2006}
\bibinfo{author}{\bibfnamefont{J.~D.} \bibnamefont{Stevenson}},
  \bibinfo{author}{\bibfnamefont{J.}~\bibnamefont{Schmalian}},
  \bibnamefont{and} \bibinfo{author}{\bibfnamefont{P.~G.}
  \bibnamefont{Wolynes}}, \bibinfo{journal}{Nature Physics}
  \textbf{\bibinfo{volume}{2}}, \bibinfo{pages}{268} (\bibinfo{year}{2006}).

\bibitem[{\citenamefont{Donati et~al.}(1998)\citenamefont{Donati, Douglas, Kob,
  Plimpton, Poole, and Glotzer}}]{Donati1998}
\bibinfo{author}{\bibfnamefont{C.}~\bibnamefont{Donati}},
  \bibinfo{author}{\bibfnamefont{J.~F.} \bibnamefont{Douglas}},
  \bibinfo{author}{\bibfnamefont{W.}~\bibnamefont{Kob}},
  \bibinfo{author}{\bibfnamefont{S.~J.} \bibnamefont{Plimpton}},
  \bibinfo{author}{\bibfnamefont{P.~H.} \bibnamefont{Poole}}, \bibnamefont{and}
  \bibinfo{author}{\bibfnamefont{S.~C.} \bibnamefont{Glotzer}},
  \bibinfo{journal}{Physical Review Letters} \textbf{\bibinfo{volume}{80}},
  \bibinfo{pages}{2338} (\bibinfo{year}{1998}).

\bibitem[{\citenamefont{Pal et~al.}(2008)\citenamefont{Pal, O'Hern,
  Blawzdziewicz, Dufresne, and Stinchcombe}}]{Pal2008}
\bibinfo{author}{\bibfnamefont{P.}~\bibnamefont{Pal}},
  \bibinfo{author}{\bibfnamefont{C.~S.} \bibnamefont{O'Hern}},
  \bibinfo{author}{\bibfnamefont{J.}~\bibnamefont{Blawzdziewicz}},
  \bibinfo{author}{\bibfnamefont{E.~R.} \bibnamefont{Dufresne}},
  \bibnamefont{and}
  \bibinfo{author}{\bibfnamefont{R.}~\bibnamefont{Stinchcombe}},
  \bibinfo{journal}{Physical Review E} \textbf{\bibinfo{volume}{78}},
  \bibinfo{pages}{011111} (\bibinfo{year}{2008}).

\bibitem[{\citenamefont{Pazmino~Betancourt
  et~al.}(2014)\citenamefont{Pazmino~Betancourt, Douglas, and
  Starr}}]{Betancourt2014}
\bibinfo{author}{\bibfnamefont{B.~A.} \bibnamefont{Pazmino~Betancourt}},
  \bibinfo{author}{\bibfnamefont{J.~F.} \bibnamefont{Douglas}},
  \bibnamefont{and} \bibinfo{author}{\bibfnamefont{F.~W.} \bibnamefont{Starr}},
  \bibinfo{journal}{The Journal of Chemical Physics}
  \textbf{\bibinfo{volume}{140}}, \bibinfo{pages}{204509}
  (\bibinfo{year}{2014}).

\bibitem[{\citenamefont{Keys et~al.}(2007)\citenamefont{Keys, Abate, Glotzer,
  and Durian}}]{Keys2007}
\bibinfo{author}{\bibfnamefont{A.~S.} \bibnamefont{Keys}},
  \bibinfo{author}{\bibfnamefont{A.~R.} \bibnamefont{Abate}},
  \bibinfo{author}{\bibfnamefont{S.~C.} \bibnamefont{Glotzer}},
  \bibnamefont{and} \bibinfo{author}{\bibfnamefont{D.~J.}
  \bibnamefont{Durian}}, \bibinfo{journal}{Nature Physics}
  \textbf{\bibinfo{volume}{3}}, \bibinfo{pages}{260} (\bibinfo{year}{2007}).
  
  \bibitem[{\citenamefont{Zhang et~al.}(2011)\citenamefont{Zhang, Yunker, Habdas,
  and Yodh}}]{Zhang2011b}
\bibinfo{author}{\bibfnamefont{Z.} \bibnamefont{Zhang}},
  \bibinfo{author}{\bibfnamefont{P.~J.} \bibnamefont{Yunker}},
  \bibinfo{author}{\bibfnamefont{P.} \bibnamefont{Habdas}},
  \bibnamefont{and} \bibinfo{author}{\bibfnamefont{A.~G.}
  \bibnamefont{Yodh}}, \bibinfo{journal}{Physical Review Letters}
  \textbf{\bibinfo{volume}{107}}, \bibinfo{pages}{208303} (\bibinfo{year}{2011}).
  
   \bibitem[{\citenamefont{Schoenholz et~al.}(2016)\citenamefont{Schoenholz, Cubuk, Sussman, Kaxiras and Liu}}]{Schoenholz2016}
  \bibinfo{author}{\bibfnamefont{S.~S.} \bibnamefont{Schoenholz}},
\bibinfo{author}{\bibfnamefont{E.~D.} \bibnamefont{Cubuk}},
  \bibinfo{author}{\bibfnamefont{D.~M.} \bibnamefont{Sussman}},
  \bibinfo{author}{\bibfnamefont{E.} \bibnamefont{Kaxiras}},
  \bibnamefont{and} \bibinfo{author}{\bibfnamefont{A.~J.}
  \bibnamefont{Liu}}, \bibinfo{journal}{Nature Physics}
  \textbf{\bibinfo{volume}{12}}, \bibinfo{pages}{469} (\bibinfo{year}{2016}).

   \bibitem[{\citenamefont{Zhao et~al.}(2013)\citenamefont{Zhao, Simon and McKenna}}]{Zhao2013}

  \bibinfo{author}{\bibfnamefont{J.} \bibnamefont{Zhao}},
  \bibinfo{author}{\bibfnamefont{S.~L.} \bibnamefont{Simon}},
  \bibnamefont{and} \bibinfo{author}{\bibfnamefont{G. B.}
  \bibnamefont{McKenna}}, \bibinfo{journal}{Nature Communications}
  \textbf{\bibinfo{volume}{4}}, \bibinfo{pages}{1783} (\bibinfo{year}{2013}).

\bibitem[{\citenamefont{Redner}(2001)}]{Redner2001}
\bibinfo{author}{\bibfnamefont{S.}~\bibnamefont{Redner}},
  \emph{\bibinfo{title}{A guide to first-passage processes}}
  (\bibinfo{publisher}{Cambridge University Press}, \bibinfo{year}{2001}).

\bibitem[{\citenamefont{Kent}(1978)}]{Kent1978}
\bibinfo{author}{\bibfnamefont{J.}~\bibnamefont{Kent}}, \bibinfo{journal}{The
  Annals of Probability} p. \bibinfo{pages}{760} (\bibinfo{year}{1978}).

\bibitem[{\citenamefont{Arfken}(1985)}]{Arfken1985}
\bibinfo{author}{\bibfnamefont{G.}~\bibnamefont{Arfken}},
  \emph{\bibinfo{title}{Inverse Laplace Transformation, in Mathematical Methods
  for Physicists}} (\bibinfo{publisher}{Academic Press, Orlando, FL},
  \bibinfo{year}{1985}).

\bibitem[{\citenamefont{Rubinstein and Colby}(2003)}]{Rubinstein2003}
\bibinfo{author}{\bibfnamefont{M.}~\bibnamefont{Rubinstein}} \bibnamefont{and}
  \bibinfo{author}{\bibfnamefont{R.~H.} \bibnamefont{Colby}},
  \emph{\bibinfo{title}{Polymer Physics}} (\bibinfo{publisher}{Oxford
  University Press}, \bibinfo{year}{2003}).

\bibitem[{\citenamefont{Kahle et~al.}(1997)\citenamefont{Kahle, Korus, Hempel,
  Unger, H\"oring, Schr\"otter, and Donth}}]{Kahle1997}
\bibinfo{author}{\bibfnamefont{S.}~\bibnamefont{Kahle}},
  \bibinfo{author}{\bibfnamefont{J.}~\bibnamefont{Korus}},
  \bibinfo{author}{\bibfnamefont{E.}~\bibnamefont{Hempel}},
  \bibinfo{author}{\bibfnamefont{R.}~\bibnamefont{Unger}},
  \bibinfo{author}{\bibfnamefont{S.}~\bibnamefont{H\"oring}},
  \bibinfo{author}{\bibfnamefont{K.}~\bibnamefont{Schr\"otter}},
  \bibnamefont{and} \bibinfo{author}{\bibfnamefont{E.}~\bibnamefont{Donth}},
  \bibinfo{journal}{Macromolecules} \textbf{\bibinfo{volume}{30}},
  \bibinfo{pages}{7214} (\bibinfo{year}{1997}).

\end{thebibliography}
\end{document}